\documentclass{mem}
\usepackage{natbib}\usepackage{txfonts}\usepackage{balance}
\usepackage{graphicx}
        \usepackage{epstopdf}
\usepackage[a4paper]{hyperref}
\idline{75}{282}

\begin{document}

\title{Evidence for a Universal Slope of the Period-Luminosity Relation
from Direct Distances to Cepheids in the LMC}

   \subtitle{}
   
\author{Wolfgang Gieren\inst{1}, 
Jesper Storm\inst{2}, 
Thomas G. Barnes III\inst{3},
Pascal Fouqu\'e\inst{4},\\
Grzegorz Pietrzy\'nski\inst{5}
\and
Francesco Kienzle\inst{6}}

\offprints{W. Gieren}
 
\institute{
Universidad de Concepci\'on, Departamento de F\'isica, Casilla 160-C, Concepci\'on, Chile
\and
Astrophysikalisches Institut Potsdam, An der Sternwarte 16, D-14482 Potsdam, Germany
\and
The University of Texas at Austin, McDonald Observatory, 1 University Station,
C1402, Austin, TX, USA 78712--0259
\and
Observatoire Midi-Pyr\'en\'ees, Laboratoire d'Astrophysique (UMR 5572),  14, avenue Edouard Belin, F-31400 Toulouse, France
\and
Universidad de Concepci\'on, Departamento de F\'isica, Casilla 160-C, Concepci\'on, Chile
\and
Observatoire de Geneve, 51 Chemin des Maillettes, CH-1290 Sauverny, Switzerland
\email{wgieren@astro-udec.cl}
}

\authorrunning{Gieren et al. }

\titlerunning{Direct Distances to Cepheids in LMC}

\abstract{We have applied the infrared surface brightness (ISB) technique to
derive distances to 13 Cepheid variables in the LMC which have periods  from
3-42 days. The corresponding absolute magnitudes define PL relations in VIWJK
bands which agree exceedingly well with the corresponding Milky Way relations
obtained from the same technique, and are in significant disagreement with the
observed LMC Cepheid PL relations, by OGLE-II and Persson et al., in these
bands. Our data uncover a systematic error in the p-factor law which
transforms Cepheid radial velocities into pulsational velocities. We correct
the p-factor law by requiring that all LMC Cepheids share the same
distance. Re-calculating all Milky Way and LMC Cepheid distances with the
revised p-factor law, we find that the PL relations from the ISB technique
both in LMC and in the Milky Way agree with the OGLE-II and Persson et al. LMC
PL relations, supporting the conclusion of no metallicity effect on the
{\em slope} of the Cepheid PL relation in optical/near infrared bands.
 } \maketitle{}

\section{Introduction}

Cepheid variables are key objects for the calibration of the extragalactic
distance scale. Using the period-luminosity (PL) relation as a tool, Cepheid
distances to galaxies out to about 25 Mpc can currently be measured (Freedman
et al. 2001).  These measurements rest on the assumption that the PL relation
is universal, or that a possible effect of metallicity on the slope/zero point
of the PL relation can be well controlled.

An alternative way to measure the distances to individual Cepheids is offered
by the infrared surface brightness (ISB) technique. This method uses the
light-, colour- and radial velocity-curves of Cepheids to calculate their
distances and mean radii. It is nearly independent of errors in reddening, and
can be applied on any individual Cepheid for which the necessary datasets have
been obtained. Very importantly, the relation between surface brightness and
colour has recently been put on a solid foundation by means of interferometric
angular diameters of Cepheids (Kervella et al. 2004a, Nordgren et al. 2002).

A disturbing issue over recent years has been the fact the the slope of the PL
relation derived from the ISB method for Milky Way Cepheids (Storm et
al. 2004,  Gieren et al. 1998) has turned out to be {\em steeper} than the
slope determined from LMC Cepheids by the various microlensing projects,
particularly by the OGLE-II project (Udalski et al. 1999), hinting at a
possible metallicity effect on the {\em slope} of the PL relation. One way to
check on this intruiging possibility, which would imply a strong complication
in the use of Cepheids as standard candles, is to measure the distances to a
number of LMC Cepheids directly with the ISB technique, and compare the
resulting LMC PL relation to the results from the OGLE-II project, and to the
results of Persson et al.  (2004) who have established accurate Cepheid PL
relations in the LMC in the near-infrared JHK bands.

\section{Infrared Surface Brightness Distances to LMC Cepheids }

We have used the ISB technique to determine the distances to 13 LMC Cepheids
with periods in the range 3-42 days. Details on the stars, and on the
different datasets used in these calculations can be found in Gieren et
al. (2005a). A detailed description of the ISB technique can be found in Storm
et al. (2004), and in the paper of Barnes et al. in these proceedings.

When plotting the absolute magnitudes of the LMC Cepheids derived from their
ISB distances against period, it turns out that in all bands from V through K,
the sequences are exceedingly well described by slopes which are identical to
the Milky Way counterpart relations, and in significant disgreement with the
slopes found by the OGLE-II project in the V, I and the reddening-free
Wesenheit (W) bands, and by Persson et al. (2004) in the near-IR J and K
bands. This is shown for the W band in Fig. 1. When plotting the individual,
tilt-corrected true LMC Cepheid distance moduli against period, a significant
trend of the moduli with period is apparent (Fig. 2). We interpret this
unphysical trend as a clear sign for the presence of a systematic,
period-dependent error in the ISB method.

\begin{figure}[]
\resizebox{\hsize}{!}{\includegraphics[clip=true]{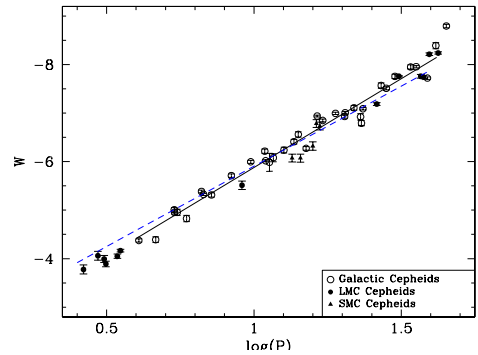}}
\caption{
\footnotesize
Absolute W-band magnitudes derived from the ISB distances of 38 Milky Way, 13
LMC and 5 SMC Cepheids. The canonical p-factor relation was used for the ISB
distance calculation for all stars. The solid line is the best fit to the
Milky Way data. The dashed line is the W-band PL relation in the LMC from the
OGLE-II project, for an assumed LMC distance modulus of 18.50. The ISB-based
PL relations both in Milky Way and LMC are significantly steeper than the
OGLE-II LMC relation.
}
\label{fig1}
\end{figure}

There are two sources of systematic error which can in principle
introduce such a period-related effect on the distances. The first is a wrong
calibration of the surface brightness relation used in the ISB technique; the
other one is a period dependence of the p-factor law which is different from
the one we assumed. Regarding the possibility of a systematic error in the
surface brightness relation, the recent  interferometric work of various
groups has nailed down the surface brightness-colour  relation to an accuracy
of $2\%$, and has confirmed the Fouqu\'e \& Gieren (1997) relation used in our
work at this level of precision. A direct comparison of the
interferometrically determined angular diameter variation of the nearby
Cepheid $\ell$~Car with that prediced by the ISB method has yielded agreement at
the $1\%$ level (Kervella et al., 2004b).  Any remaining systematic
uncertainty on the surface brightness-colour relation for Cepheids can
therefore not explain the trend observed in Fig. 2.

\begin{figure}[]
\resizebox{\hsize}{!}{\includegraphics[clip=true]{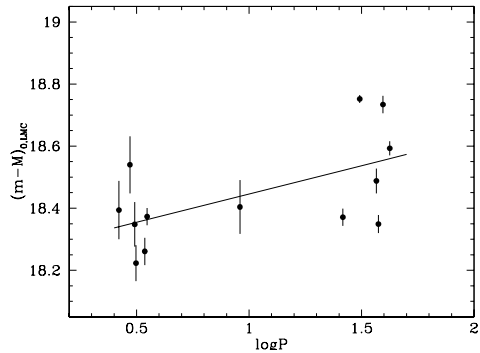}}
\caption{
\footnotesize
ISB-determined true distance moduli for LMC Cepheids, calculated with the
canonical p-factor law. The distance moduli have been corrected for the tilt
of the LMC plane with respect to the line of sight. There is a significant
trend of the distance moduli with period.
}
\label{fig2}
\end{figure}

Regarding the p-factor law, we have assumed p=1.39 - 0.03 logP (P in days), as
established by Gieren et al. (1993) from the theoretical models of Hindsley \&
Bell (1986). Our current work suggests that the assumed period dependence in
this law is seriously flawed. We have therefore re-determined the period
dependence of the p-factor law by requiring the trend seen in Fig. 2 to
disappear. Since we cannot be confident of the zero point of the adopted
p-factor law either, in view of our new results for the LMC Cepheid distances,
we chose to use twelve well-established open cluster Cepheids in the Milky Way
which have both ZAMS-fitting and ISB distance determinations to set the zero
point of the p-factor law, by the requirement that the average difference
between the ZAMS-fitting and ISB distances be zero for this sample of stars
(see Gieren et al. 2005a for details). Actually, the derivation of the slope
and zero point of the new p-factor law are not completely independent from
each other.  Our result for the {\em revised p-factor law} is p=1.58 ($\pm$
0.02) - 0.15 ($\pm$ 0.05) log P, suggesting that the p factor to be used in
Baade-Wesselink work on Cepheid variables  depends more strongly on period
(and thus luminosity) than previously thought. The expression for p is also
tied to an assumed LMC distance of 18.56 mag.

\begin{figure}[]
\resizebox{\hsize}{!}{\includegraphics[clip=true]{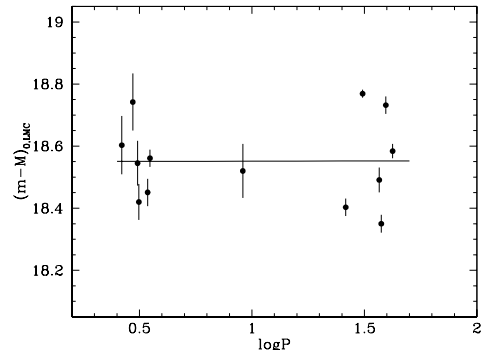}}
\caption{
\footnotesize
As Fig. 2, but now distance moduli calculated with the revised p-factor law of
this paper. The trend in Fig. 2 has disappeared.
}
\label{fig3}
\end{figure}

We have re-calculated all ISB distances (LMC Cepheids, and Milky Way Cepheids)
with the revised p-factor law. Fig. 3 displays the corrected LMC Cepheid
distance moduli, plotted against the periods of the stars - evidently, there
is no trend anymore. When we re-calculate the slopes of the PL relations in
the different bands, we find that the excellent agreement between Milky Way
and LMC relations persists, but now these relations are all shallower than
before, and agree with the OGLE-II and Persson et al. PL relations {\em within
the combined $1\sigma$ uncertainties}. This is demonstrated in Fig. 4 for the
K band.

\begin{figure}[]
\resizebox{\hsize}{!}{\includegraphics[clip=true]{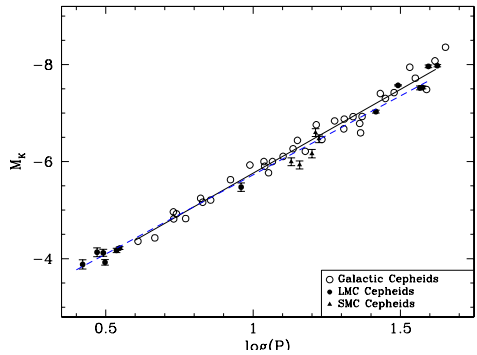}}
\caption{
\footnotesize
Absolute ISB-based K-band PL relations in Milky Way and LMC, calculated 
from the revised p-factor law. Solid line is the best fit to the
Milky Way data. This line fits also the LMC data very well. The
dashed line is the Persson et al. PL relation for LMC Cepheids, which
now agrees very well with the ISB-based Milky Way and LMC PL relations.
}
\label{fig4}
\end{figure}

\section{Conclusions}

Our current results strongly suggest that the previous problem with Milky Way
Cepheid PL relations from the ISB technique being steeper than LMC and SMC
relations was caused by an erroneous calibration of the p-factor law used to
convert the observed Cepheid radial velocities into their pulsational
velocities. With the corrected law, the discrepancy disappears, and we now get
Milky Way PL relations from the ISB technique  in all bands which agree with
the PL relations observed in the corresponding bands in the LMC by the
microlensing projects, and by Persson et al. in the near-infrared. Adding to
this excellent agreement between Milky Way and LMC our current knowledge about
the PL relations in more metal-poor galaxies, like IC 1613 (Udalski et
al. 2001), there seems now solid evidence that the {\em slope} of the Cepheid
PL relation in the optical VIW bands is independent of metallicity, at least
in the metallicity regime from -1.0 dex to solar. In order to investigate if
this is also true for near-infrared bands, more work on near-IR  PL relations
in nearby galaxies must be done. The Araucaria Project discussed by
Pietrzy{\'n}ski et al. elsewhere in these proceedings will provide such
studies in the very near future. One recent example is the near-IR Cepheid
work in NGC 300 (Gieren et al.  2005b).

In order to address remaining concerns with our present work, we need to
obtain ISB distances to a significant number of additional Cepheids in the
LMC, in order to improve the statistics of our results. Such a program has
recently been started. In particular, it will be important to analyze a
number of short-period Cepheids in the general LMC field which are not related
to each other as the ones in our current study, all which are all members of
the same cluster, NGC 1866 (see Storm et al. in these proceedings).

\begin{acknowledgements}

WG and GP acknowledge financial support for this work from the Chilean Center
for Astrophysics  FONDAP 15010003.

\end{acknowledgements}

\bibliographystyle{aa}

\end{document}